%% file: main.tex
\documentclass[10pt, conference, letterpaper]{IEEEtran}
\IEEEoverridecommandlockouts
\usepackage{epsfig}
\usepackage{color,booktabs}
\usepackage{graphicx}
\usepackage[utf8]{inputenc}
\usepackage[figuresright]{rotating}
\usepackage{amssymb}
\usepackage{amsmath}
\usepackage{setspace}
\usepackage{rotating}
\usepackage{multirow}
\usepackage{wrapfig}
\usepackage{booktabs}
\usepackage{array}
\usepackage{cite}
\usepackage{float}
\usepackage{comment}
\usepackage{pict2e}
\usepackage{epstopdf}
\usepackage[ruled]{algorithm2e}
\usepackage{xcolor}
\DeclareMathOperator*{\argmin}{\arg\!\min}

\usepackage{balance}
\usepackage{mathtools}
\usepackage{cite}
\usepackage[hyphens]{url}
\usepackage[labelformat=simple]{subcaption}
\usepackage{booktabs}
\usepackage{url}

\usepackage{enumitem}

\usepackage[font=footnotesize, tableposition=top]{caption}
\usepackage{xparse}
\usepackage[T1]{fontenc}
\input{preamble/units}

\input{preamble/markupAndCommenting}

\newcommand{\newtext}[1]{{\color{black}#1}}

\title{E2E-WAVE: End-to-End Learned Waveform Generation for Underwater Video Multicasting}

\author{{\bf Khizar Anjum*, Tingcong Jiang*, and Dario Pompili}\\
Department of Electrical and Computer Engineering, Rutgers University--New Brunswick, NJ, USA\\
\textit{\{khizar.anjum, tingcong.jiang, pompili\}@rutgers.edu}
\thanks{*Khizar Anjum and Tingcong Jiang contributed equally to this work. Open-source reproducibility repository: \url{https://github.com/khizar-anjum/e2e-wave.git}.}
}

\begin{document}

\bstctlcite{IEEEexample:BSTcontrol}

\maketitle

\IEEEpeerreviewmaketitle

\begingroup
\renewcommand\thefootnote{}
\footnotetext{\scriptsize \copyright~2026 IEEE. Personal use of this material is permitted. Permission from IEEE must be obtained for all other uses, in any current or future media, including reprinting/republishing this material for advertising or promotional purposes, creating new collective works, for resale or redistribution to servers or lists, or reuse of any copyrighted component of this work in other works. Accepted to IEEE SECON 2026.}
\endgroup

\begin{abstract}
We present E2E-WAVE, the first end-to-end learned waveform generation system for underwater video multicasting. Acoustic channels exhibit 20--46\% bit error rates where forward error correction becomes counterproductive---LDPC increases rather than decreases errors beyond its decoding threshold. E2E-WAVE addresses this by embedding semantic similarity directly into physical layer waveforms: when decoding errors are unavoidable, the system preferentially selects semantically similar tokens rather than arbitrary corruption. Combining VideoGPT tokenization (1024$\times$ compression) with a trainable waveform bank and fully differentiable OFDM transmission, E2E-WAVE achieves +5~dB (19.26\%) PSNR and +0.10 (14.28\%) SSIM over the strongest FEC-protected baseline in less challenging underwater channel (NOF1) while delivering real-time 16~FPS video at 128$\times$128 resolution over 2.3~kbps channels---impossible for conventional digital modulation. The performance gap only increases in harsher channels (BCH1, NCS1). Trained on a single channel, E2E-WAVE generalizes to unseen underwater environments without retraining, while HEVC fails at sub-5~kbps rates and SoftCast's AWGN assumptions collapse on frequency-selective channels.
\end{abstract}

\section{Introduction}\label{sec:intro}

Video transmission underwater enables critical scientific and industrial applications including marine biodiversity monitoring, infrastructure inspection, and oceanographic research~\cite{sahoo2019advancements}. However, underwater acoustic channels impose severe constraints: limited bandwidth (kHz range), long delay spreads (100+ ms), frequency-selective fading, multipath propagation, and Doppler effects~\cite{akyildiz2005underwater}. While optical and radio frequency alternatives offer higher bandwidth, they support only short-range links (1--100m)~\cite{heidemann2012underwater}, making acoustic communication the only viable option for long-range underwater networks despite effective bitrates in the kbps range.

\textbf{Prior Approaches and Limitations:}
Existing underwater video transmission approaches follow two paradigms. First, conventional systems compress video to discrete representations using standard codecs (AVC, HEVC, VVC)~\cite{ISO83529, martinez-rach_performance_2021} or learned neural video tokenizers that encode video frames as sequences of discrete indices, where each index selects a token (a learned vector code) from a finite codebook representing compressed spatiotemporal video content. Transmission is then protected with forward error correction (FEC) codes such as turbo codes or Low-Density Parity-Check~(LDPC) codes. However, these codecs are designed for terrestrial high-bandwidth environments and exhibit catastrophic failures at ultra-low bitrates typical of underwater links. Moreover, conventional channel coding treats all bit errors equally, ignoring the semantic structure of video data. When bit errors occur in transmitted indices, the decoder retrieves semantically distant tokens from the codebook, producing visually incoherent artifacts rather than graceful degradation.
Second, energy-based analog approaches such as SoftCast~\cite{jakubczak2010softcast} and ECast~\cite{zhang2015ecast} transmit uncoded transform coefficients, emphasizing metadata protection to enable receivers to decode based on received signal energy. While this provides graceful SNR scaling in AWGN channels, underwater channels exhibit severe frequency-selective fading and multipath that dramatically alter the energy distribution across frequency bins. Furthermore, the additionally transmitted metadata consumes bandwidth and any corruption in the metadata will introduces catastrophic failures, making the systems vulnerable.

\textbf{Our Approach:}
We propose E2E-WAVE (End-to-End Waveform Adaptive Encoding), motivated by a key intuition: if decoding errors are unavoidable under harsh channel conditions, the system should preferentially decode to semantically similar tokens rather than arbitrary incorrect tokens. To achieve this, we embed semantic similarity structure directly into the physical layer waveforms, enabling control over decoding errors at the finest granularity. Unlike prior work~\cite{abdi2025phydnns} that maps video features to predefined digital modulations (BPSK, QPSK), E2E-WAVE is the first system to learn transmission waveforms end-to-end using a data-driven approach. We construct a trainable waveform bank that maps each discrete video token to a learned complex-valued waveform, optimized such that semantically similar tokens correspond to waveforms with small Euclidean distances in the signal space. This waveform bank controls the distribution of decoding errors: when channel noise causes token misdetection, the nearest-neighbor search preferentially selects semantically similar tokens, minimizing perceptual distortion.

To enable end-to-end training, we develop a fully differentiable channel simulation pipeline that supports OFDM equalization with pilot-based channel estimation and replays channels collected from field experiements to simulate channel effects, including frequency-selective fading, multipath, Doppler shifts. Gradients backpropagate from cross-entropy loss through the entire transmission stack to waveform parameters, jointly optimizing for both channel robustness and semantic preservation. Our system achieves real-time video transmission at 16~FPS with 128$\times$128 resolution over severely bandwidth-constrained underwater acoustic channels, outperforming conventional digital modulation with FEC and analog SoftCast baselines by substantial margins in Peak Signal-to-Noise Ratio~(PSNR) and Structural Similarity Index Measure~(SSIM) metrics across diverse channel conditions.

\begin{figure*}
    \centering
    \includegraphics[width=\linewidth]{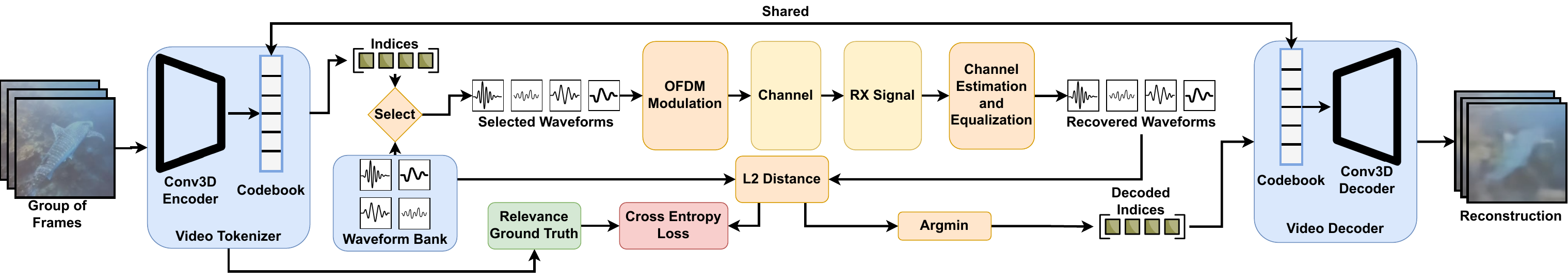}
    \caption{\textbf{TX}: Conv3D encoder compresses video frames to token indices (1024x) via shared codebook. Indices select learned waveforms from a trainable bank, undergo OFDM modulation (preamble, pilots), and transmit through collected at-sea channels with fading, multipath, and Doppler. \textbf{RX}: Synchronization, pilot-based estimation, and zero-forcing equalization recover waveforms. L2-based argmin decodes token indices, retrieving codebook entries for Conv3D reconstruction. Cross-entropy loss between decoded token distribution (from L2 distances) and semantic relevance (codebook similarity) backpropagates through the differentiable pipeline to embed semantic structure into physical layer waveforms. Encoder, decoder, and codebook are pretrained and kept frozen.}
    \label{fig:e2e-wave-architecture}
    \vspace{-0.4cm}
\end{figure*}

\textbf{Our Contributions:}
\begin{itemize}
    \item To the best of our knowledge, E2E-WAVE is the first one to embed semantic similarity directly into physical layer waveforms. Our proposed waveform bank is optimized via cross-entropy loss against semantic relevance, ensuring channel-induced decoding errors preferentially select semantically similar tokens for graceful degradation.

    \item We develop an end-to-end differentiable transmission simulation pipeline supporting OFDM with pilot-based equalization. Differentiable resampling (linear interpolation) and channel replay (matrix convolutions with polyphase filtering) enable gradients to backpropagate through the entire stack, jointly optimizing waveforms for channel robustness and semantic preservation.

    \item Trained exclusively on one channel, E2E-WAVE generalizes to unseen channels without retraining or metadata, demonstrating practical deployability.

    \item E2E-WAVE achieves 16~FPS at 128$\times$128 resolution over underwater channels without metadata, outperforming digital modulation with FEC, HEVC, and SoftCast by substantial margins in PSNR/SSIM.
\end{itemize}

\textbf{Paper Outline:}
In Sect.~\ref{sec:relwork}, we position our work with respect to related literature. In Sect.~\ref{sec:solution}, we detail our proposed E2E-WAVE architecture. In Sect.~\ref{sec:perfeval}, we evaluate our approach against baseline methods across realistic underwater acoustic channels. In Sect.~\ref{sec:concl}, we conclude and discuss future work.

\section{Related Work}\label{sec:relwork}
We organized related work into three areas: video coding, transmission paradigms (covering both digital modulation and soft analog delivery), and underwater video transmission.

\textbf{Advances in Video Coding:}
Recent video coding standards---High Efficiency Video Coding~(HEVC), Joint Exploration Model~(JEM), and Versatile Video Coding~(VVC)---achieve successive 50\% bit rate reductions at equivalent quality, with VVC halving HEVC's rate, which itself halved AVC's~\cite{ISO83529}. However, these gains come at the cost of increased computational complexity~\cite{martinez-rach_performance_2021}, which is prohibitive for resource-constrained underwater platforms. Moreover, these codecs remain brittle at ultra-low bitrates (kbps range) typical of underwater acoustic links, where packet losses cause catastrophic decoding failures.
An alternative approach is neural video tokenization, which maps frames to discrete tokens from a learned codebook. Recent tokenizers such as Cosmos~\cite{nvidia2025cosmosworldfoundationmodel} achieve high compression (2048$\times$) but require minutes per group of frames on high-end GPUs, while OmniTokenizer~\cite{wang2024omnitokenizerjointimagevideotokenizer} is fast but provides insufficient compression (64$\times$) with limited reconstruction quality (~16~dB PSNR). We adopt VideoGPT's VQ-VAE~\cite{yan2021videogpt}, which achieves 1024$\times$ compression at $\sim$24~dB PSNR with real-time encoding---an 8~dB gain over OmniTokenizer at 16$\times$ higher compression.

\textbf{Transmission Paradigms:}
Conventional systems separate source and channel coding, mapping encoded bits to fixed modulation schemes (PSK, QAM, OFDM). Although effective for terrestrial links with high SNR, they struggle with underwater acoustic channels characterized by multipath fading, Doppler shifts, and ambient noise variations. Recent learned communication systems~\cite{xie2021deep, oshea2017introduction} jointly optimize transmitter-receiver chains but focus on short message transmission rather than video.
An alternative paradigm is soft delivery, introduced by SoftCast~\cite{jakubczak2010softcast}, which transmits analog coefficients directly, allowing received quality to scale with SNR. Extensions such as ECast~\cite{zhang2015ecast} address bandwidth-limited environments but require channel state feedback from the receiver for power allocation. Our E2E-WAVE approach operates without channel feedback, making it suitable for multicast scenarios where feedback from multiple receivers is impractical.

\textbf{Underwater Video Transmission:}
Among underwater transmission media (acoustic, optical, RF)~\cite{furqan2019underwater}, only acoustic waves support long-range communication (up to 20~km) independent of water turbidity. However, acoustic channels are severely bandwidth-limited (kHz) with low propagation speed (1500~m/s) and depth-dependent non-linearities~\cite{akyildiz2005underwater}. While some works address learned image compression for low-bitrate links~\cite{anjum2022deep, 9150712}, video transmission research has focused on increasing physical-layer bitrates to support standard codecs~\cite{10102549} rather than adapting the coding algorithms themselves.
E2E-WAVE addresses these gaps by jointly learning video compression and waveform generation end-to-end. Unlike standard codecs, it operates within underwater bandwidth constraints ($\sim$2.3 kbps) and delivers 16 FPS video.

\section{Proposed E2E-WAVE Architecture}\label{sec:solution}

\newtext{Compression of video into discrete semantic tokens is a well-known idea. However, our approach stems from the observation that using conventional modulation and FEC to communicate such semantic content is counter-productive, i.e., they treat all bit errors equally, ignoring semantic relationships between semantic tokens---a single-bit error can decode to a semantically distant token, causing catastrophic degradation. E2E-WAVE embeds semantic similarity into physical layer waveforms, ensuring errors preferentially map to similar tokens. We present three components: video tokenization establishing semantic relationships, a trainable waveform bank preserving semantic structure, and an OFDM pipeline for frequency-selective channels (Fig.~\ref{fig:e2e-wave-architecture}).}

\subsection{Video Tokenization}

\newtext{The E2E-WAVE system architecture comprises three primary stages: (1)~video tokenization, (2)~learned waveform transmission, and (3)~video reconstruction. At the transmission side, an input video sequence is first compressed into a sequence of discrete tokens $\mathbf{t} = [t_1, t_2, \ldots, t_N]$ through a learned video tokenizer, where each token $t_i \in \{0, 1, \ldots, K-1\}$ represents a compact encoding of spatiotemporal video information. The tokenizer's codebook $\mathcal{C} = \{\mathbf{c}_0, \mathbf{c}_1, \ldots, \mathbf{c}_{K-1}\}$ defines semantic relationships between tokens through pairwise $L_2$ distances $d(t_i, t_j) = \|\mathbf{c}_i - \mathbf{c}_j\|_2$. We construct a semantic relevance matrix $\mathbf{R} \in [0,1]^{K \times K}$ by first computing all pairwise distances $D_{ij} = d(t_i, t_j)$, applying min-max normalization to obtain $\tilde{D}_{ij} = \frac{D_{ij} - \min(D)}{\max(D) - \min(D)}$, and defining relevance as $R_{ij} = 1 - \tilde{D}_{ij}$. This relevance matrix captures the semantic similarity structure: $R_{ij} \approx 1$ indicates tokens $t_i$ and $t_j$ correspond to visually similar content, such that decoding errors between them produce perceptually coherent artifacts rather than catastrophic distortions.}

\newtext{For underwater deployment, the tokenizer must balance compression ratio, reconstruction quality, and computational efficiency. Among the candidates reviewed in Sect.~\ref{sec:relwork}, Cosmos~\cite{nvidia2025cosmosworldfoundationmodel} reaches 2048$\times$ compression but requires minutes per clip on a high-end GPU, while OmniTokenizer~\cite{wang2024omnitokenizerjointimagevideotokenizer} runs faster yet delivers only $\sim$16~dB PSNR at 64$\times$ compression---insufficient bandwidth savings for real-time underwater video even at reduced resolution. VideoGPT~\cite{yan2021videogpt} attains $\sim$24~dB PSNR at 1024$\times$ compression with real-time encoding on commodity hardware, an 8~dB gain over OmniTokenizer at 16$\times$ higher compression. We therefore adopt VideoGPT's VQ-VAE as our tokenizer backbone.}

\subsection{Learned Waveform Bank}

\newtext{Given the strong compression and reconstruction quality provided by VideoGPT tokenization, a straightforward approach for underwater video transmission would directly modulate the token indices using conventional digital schemes. Specifically, each token index $t_i \in \{0, 1, \ldots, K-1\}$ can be represented with $\log_2(K)$ bits, which are then modulated onto symbols using BPSK or QPSK. At the receiver, symbols are demodulated through hard decision decoding to recover the bit sequence, which is converted back to token indices for video reconstruction. However, this conventional approach exhibits a catastrophic cliff effect: reconstruction quality degrades precipitously with bit error rate (BER). As demonstrated in our performance evaluation, even a few bit errors (3 to 5 bits) in the token index representation significantly caps the achievable reconstruction quality, rendering video transmission extremely brittle and unreliable for practical underwater deployment. This brittleness arises because conventional modulation treats all bit errors equally, such that a single-bit error can cause the system to decode to a semantically distant token with drastically different visual content. Therefore, a more robust transmission mechanism is required that exploits the semantic similarity structure established in the tokenization stage.}

\newtext{E2E-WAVE addresses this challenge through a trainable waveform bank that directly maps each token to a learned complex acoustic waveform (Fig.~\ref{fig:e2e-wave-architecture}). To enable gradient-based optimization of complex-valued waveforms within standard deep learning frameworks, we parameterize the waveform bank in the frequency domain. Specifically, for each token $i \in \{0, 1, \ldots, K-1\}$, we maintain trainable parameters $\mathbf{F}_{\text{real}}[i, :] \in \mathbb{R}^{L}$ and $\mathbf{F}_{\text{imag}}[i, :] \in \mathbb{R}^{L}$ representing the real and imaginary components of the frequency-domain representation. The complex frequency-domain representation is constructed as $\mathbf{F}[i, :] = \mathbf{F}_{\text{real}}[i, :] + j\mathbf{F}_{\text{imag}}[i, :]$, where $j = \sqrt{-1}$. The time-domain waveform for token $i$ is obtained by applying the inverse discrete Fourier transform:}
\begin{align}
\mathbf{w}_i = \text{IDFT}(\mathbf{F}[i, :]) = \frac{1}{\sqrt{L}} \sum_{k=0}^{L-1} \mathbf{F}[i, k] \exp\left(j\frac{2\pi ik}{L}\right) \in \mathbb{C}^{L}
\end{align}
\newtext{This frequency-domain parameterization offers two advantages: (1)~it naturally handles complex-valued waveforms through separate real and imaginary components, facilitating gradient computation, and (2)~it allows direct control over the spectral characteristics of the learned waveforms.}
\newtext{The encoding process at the transmitter maps each token $t_i$ to its corresponding time-domain waveform,}
\begin{align}
\mathbf{w}_{\text{tx}} = \mathbf{w}_{t_i} = \text{IDFT}(\mathbf{F}[t_i, :]) \in \mathbb{C}^{L}
\end{align}
\newtext{Each waveform sample is transmitted as one OFDM complex data symbol; thus $L$, which we term the \textit{wavelength}, determines the number of complex data symbols required per token. Given a channel supporting $R$ complex data symbols per second, E2E-WAVE achieves a token rate of $R/L$ tokens per second. Varying $L$ trades off robustness against throughput, enabling fair comparison with FEC-protected digital baselines at equivalent rates.}

\newtext{At the receiver, after channel transmission and OFDM equalization (detailed in the next subsection), the received signal $\mathbf{r} \in \mathbb{C}^{L}$ undergoes waveform demodulation through nearest-neighbor search in the waveform bank. The decoder computes $L_2$ distances between the received waveform and all $K$ waveforms,}
\begin{align}
d_i = \|\mathbf{r} - \mathbf{w}_i\|_2, \quad i \in \{0, 1, \ldots, K-1\}
\end{align}
\newtext{and selects the token corresponding to the minimum distance:}
\begin{align}
\hat{t} = \argmin_{i \in \{0, 1, \ldots, K-1\}} d_i
\end{align}
\newtext{The recovered token $\hat{t}$ is then passed to the VideoGPT decoder to reconstruct the corresponding video frame.}

\newtext{To train the waveform bank parameters $\mathbf{F}_{\text{real}}$ and $\mathbf{F}_{\text{imag}}$ to embed the semantic relevance structure into the physical layer, we apply cross-entropy loss to the softmax of the inverse $L_2$ distances. During training, for a transmitted token $t$, we compute the predicted probability distribution over all tokens,}
\begin{align}
p_i = \frac{\exp(-d_i/\tau)}{\sum_{j=0}^{K-1} \exp(-d_j/\tau)}
\end{align}
\newtext{where $\tau > 0$ is a trainable temperature parameter controlling the sharpness of the distribution. The cross-entropy loss is then evaluated against the semantic relevance vector $\mathbf{R}_{t,:} \in [0,1]^{K}$ for the transmitted token $t$:}
\begin{align}
\mathcal{L}_{\text{wavebank}} = -\sum_{i=0}^{K-1} R_{t,i} \log(p_i)
\end{align}
\newtext{This formulation encourages the waveform bank to minimize $L_2$ distances between waveforms corresponding to tokens with high semantic relevance $R_{t,i} \approx 1$. Consequently, when channel degradation causes the receiver to select an incorrect waveform, it preferentially selects one corresponding to a semantically similar token, gracefully degrading perceptual quality rather than producing catastrophic visual artifacts.}

\begin{figure*}
	\centering
	\begin{subfigure}[b]{0.19\textwidth}
		\centering
		\includegraphics[width=\textwidth]{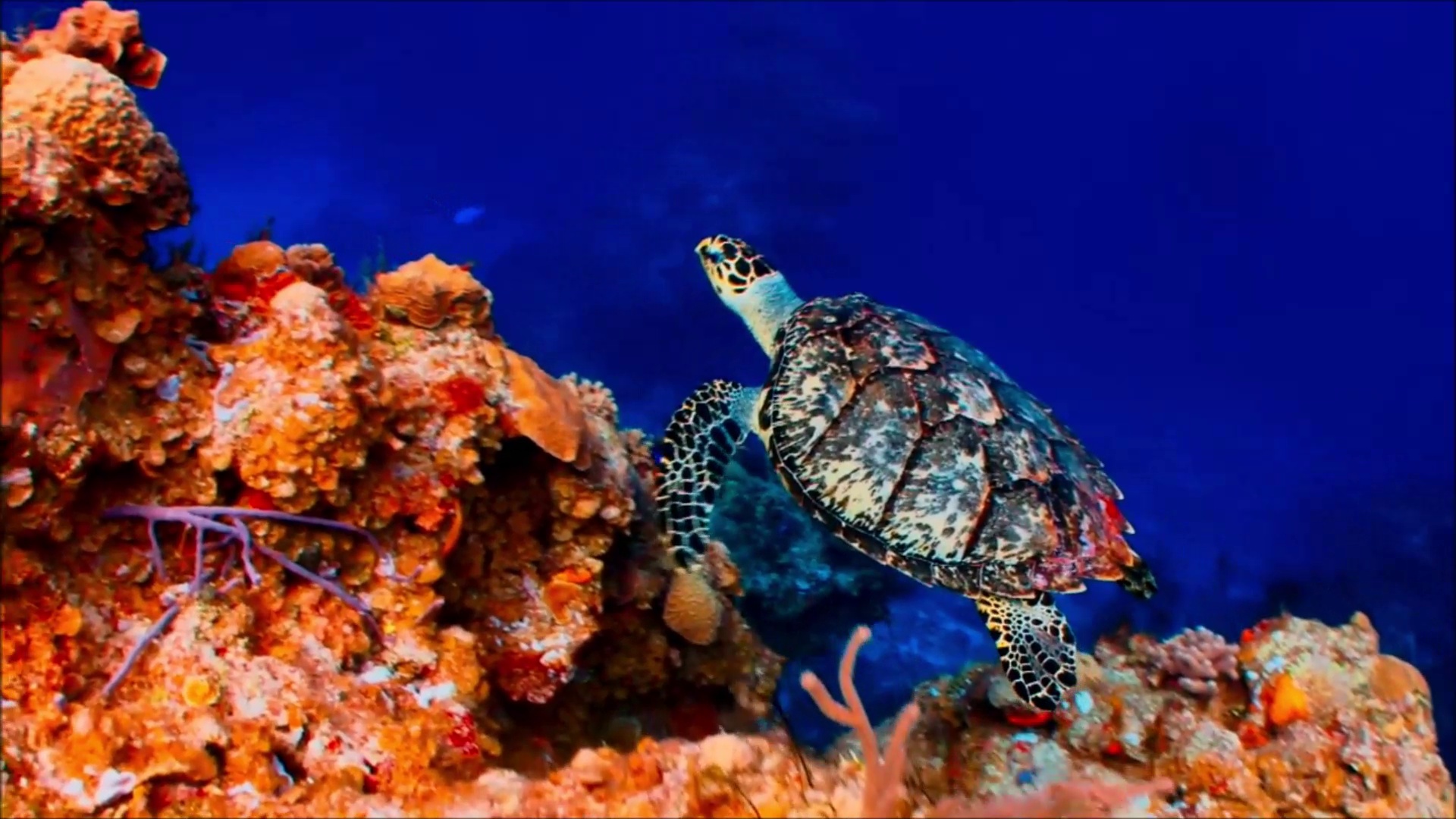}
	\end{subfigure}
	\hfill
	\begin{subfigure}[b]{0.19\textwidth}
		\centering
		\includegraphics[width=\textwidth]{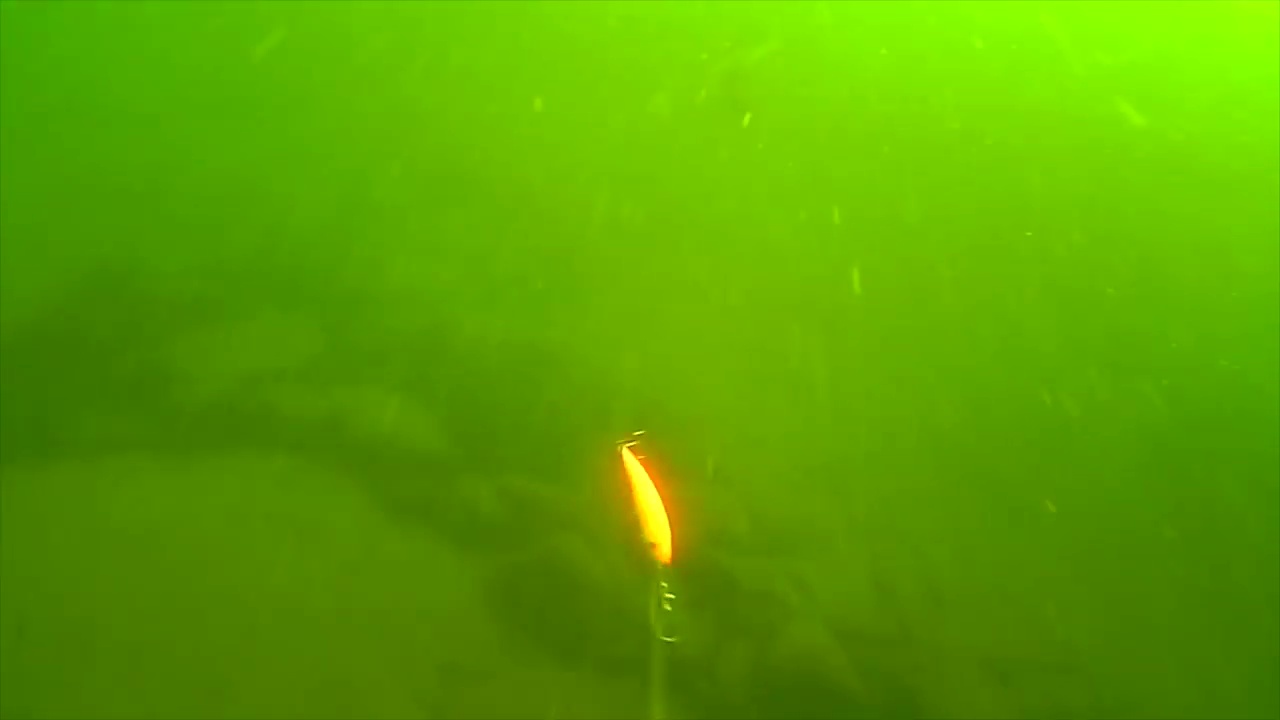}
	\end{subfigure}
	\hfill
	\begin{subfigure}[b]{0.19\textwidth}
		\centering
		\includegraphics[width=\textwidth]{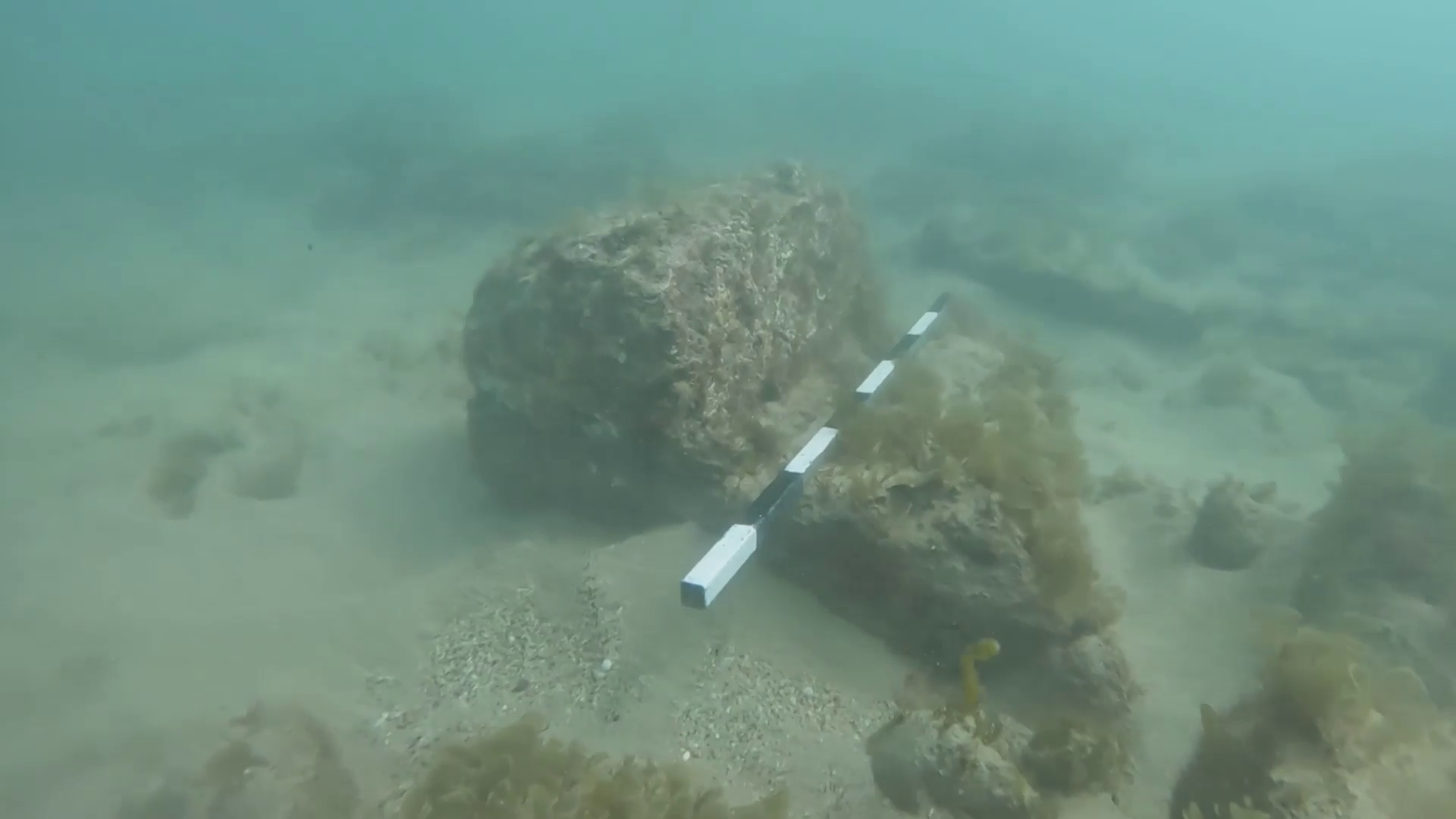}
	\end{subfigure}
	\hfill
	\begin{subfigure}[b]{0.19\textwidth}
		\centering
		\includegraphics[width=\textwidth]{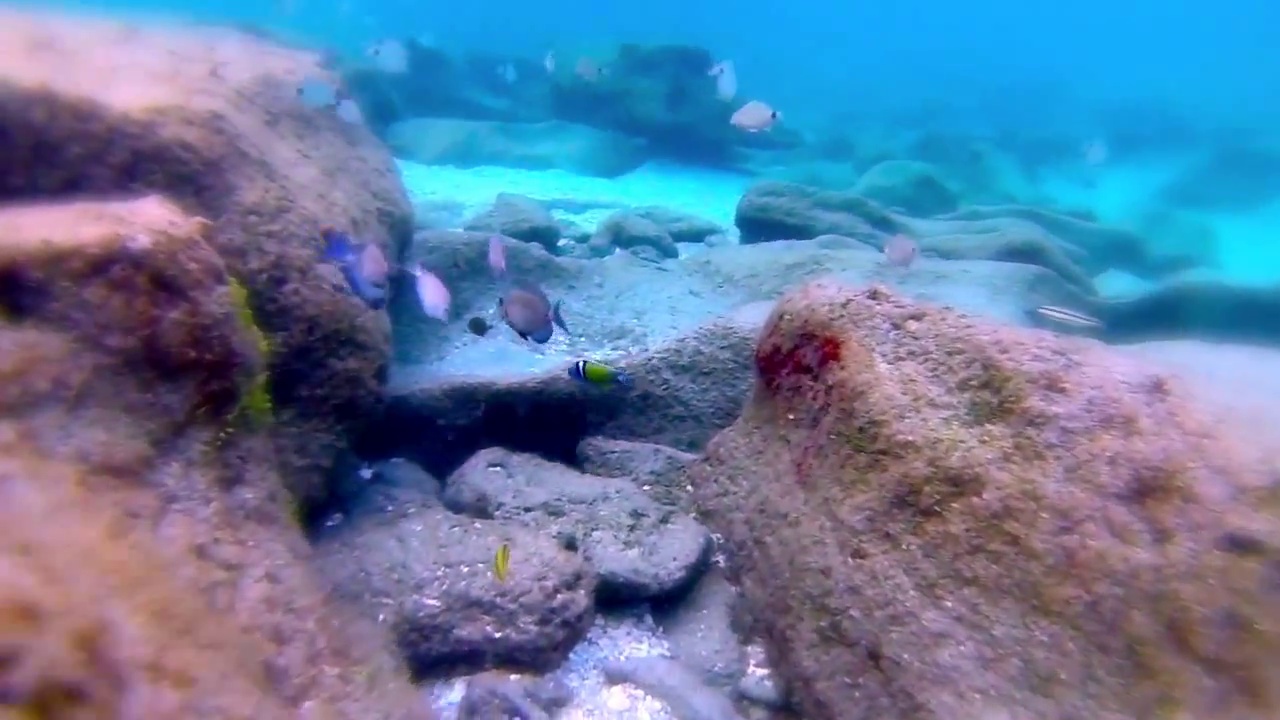}
	\end{subfigure}
	\hfill
	\begin{subfigure}[b]{0.19\textwidth}
		\centering
		\includegraphics[width=\textwidth]{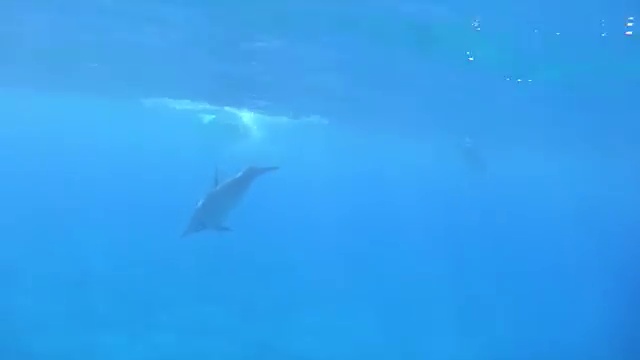}
	\end{subfigure}

	\vspace{0.5em}

	\begin{subfigure}[b]{0.115\textwidth}
		\centering
		\includegraphics[width=\textwidth]{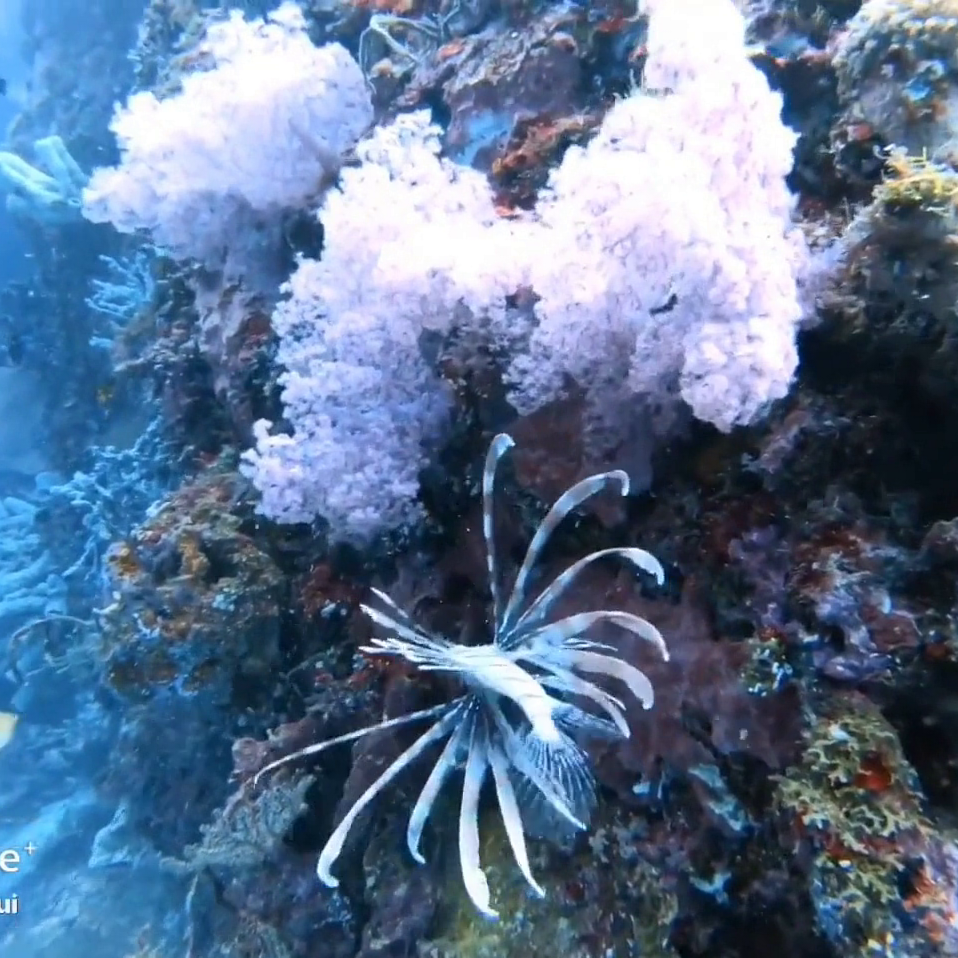}
	\end{subfigure}
	\hfill
	\begin{subfigure}[b]{0.115\textwidth}
		\centering
		\includegraphics[width=\textwidth]{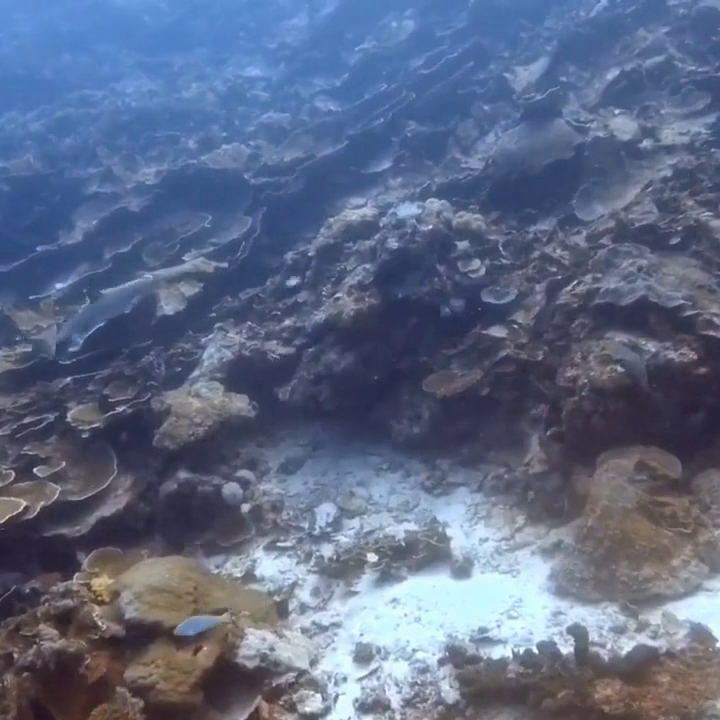}
	\end{subfigure}
	\hfill
	\begin{subfigure}[b]{0.115\textwidth}
		\centering
		\includegraphics[width=\textwidth]{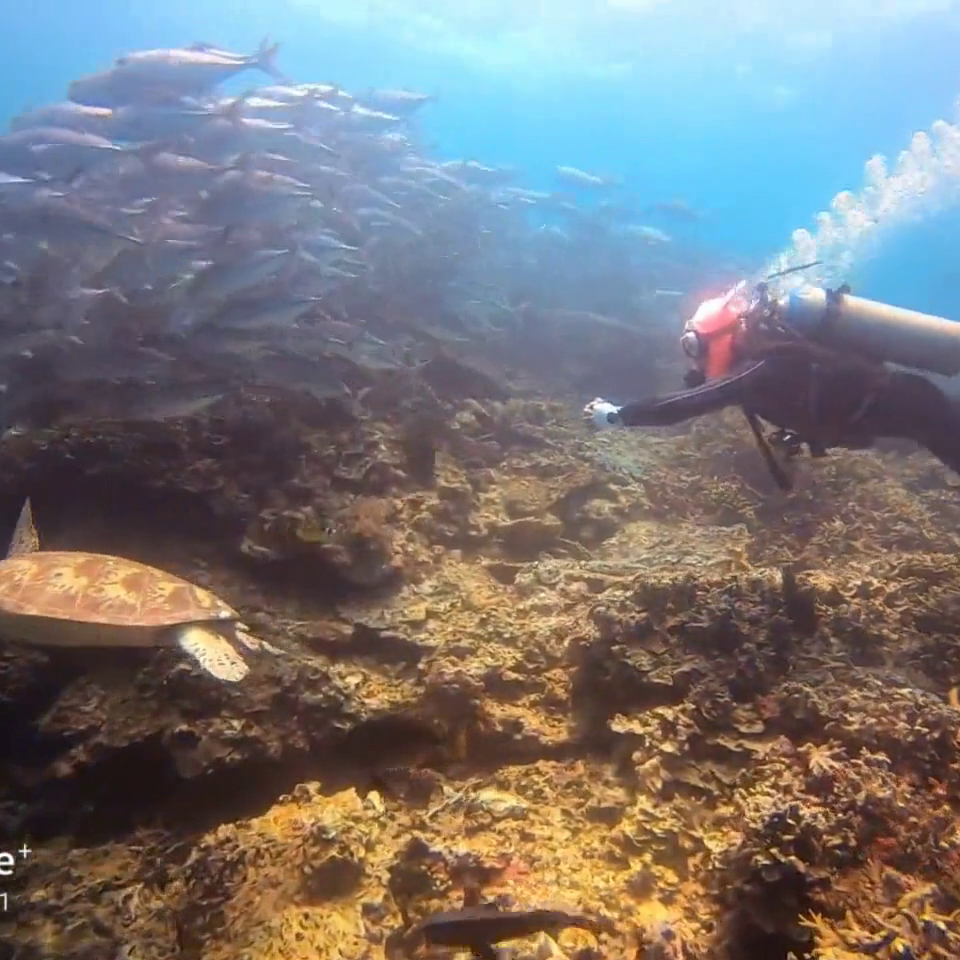}
	\end{subfigure}
	\hfill
	\begin{subfigure}[b]{0.115\textwidth}
		\centering
		\includegraphics[width=\textwidth]{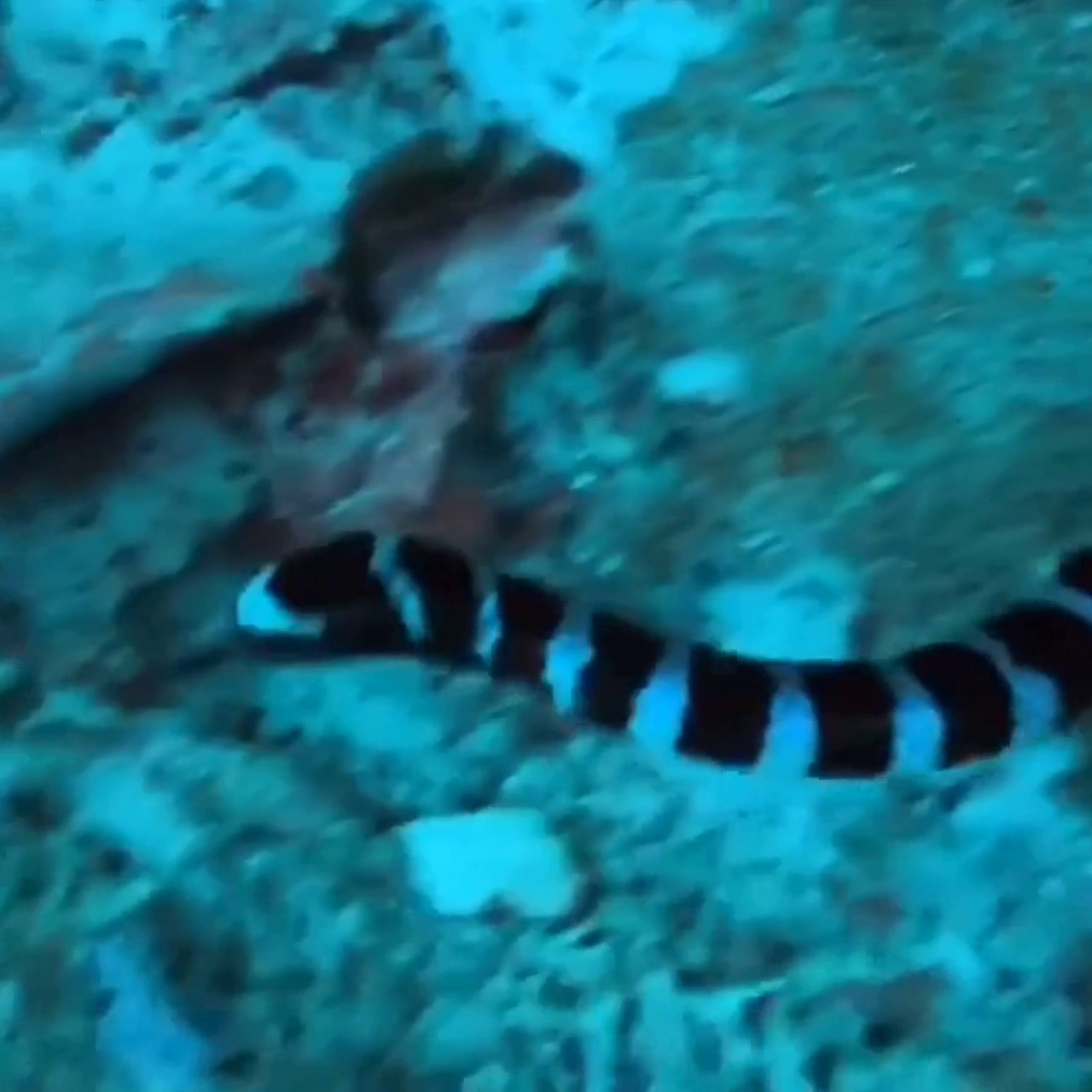}
	\end{subfigure}
	\hfill
	\begin{subfigure}[b]{0.115\textwidth}
		\centering
		\includegraphics[width=\textwidth]{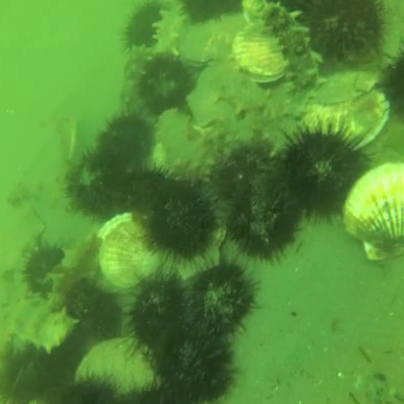}
	\end{subfigure}
	\hfill
	\begin{subfigure}[b]{0.115\textwidth}
		\centering
		\includegraphics[width=\textwidth]{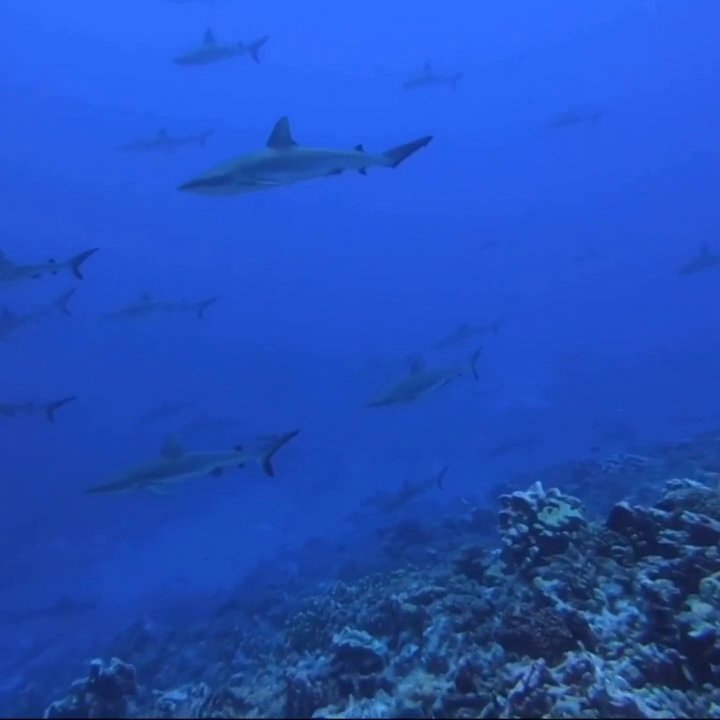}
	\end{subfigure}
	\hfill
	\begin{subfigure}[b]{0.115\textwidth}
		\centering
		\includegraphics[width=\textwidth]{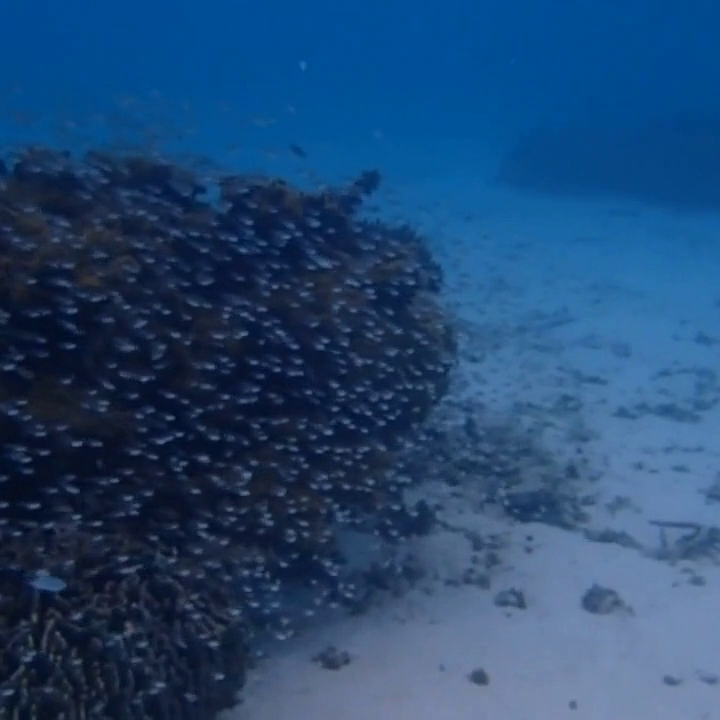}
	\end{subfigure}
	\hfill
	\begin{subfigure}[b]{0.115\textwidth}
		\centering
		\includegraphics[width=\textwidth]{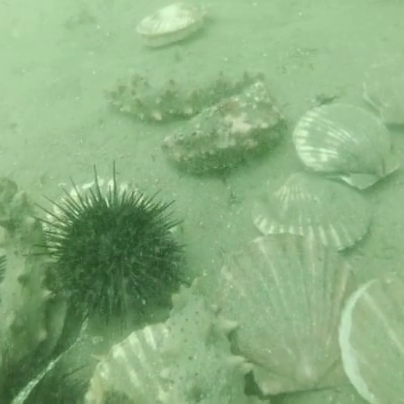}
	\end{subfigure}
	\caption{Dataset examples. \textbf{Top row}: Training data from DRUVA and UOT100 exhibiting characteristic underwater artifacts (color distortion, reduced visibility, marine snow). \textbf{Bottom row}: UVE-38K test set partitioned into \textit{clear} (left four, vivid colors and fine detail) and \textit{turbid} (right four, low visibility and diffuse lighting) categories.}
	\label{fig:dataset_examples}
    \vspace{-0.4cm}
\end{figure*}

\subsection{Transmission Pipeline}

\newtext{The learned complex waveforms $\mathbf{w}_i \in \mathbb{C}^{L}$ cannot be directly transmitted over underwater acoustic channels without explicit channel compensation. Underwater channels exhibit severe frequency-selective fading, multipath propagation with delay spreads, and Doppler effects that would dominate the received signal and prevent waveform bank convergence during training. We leverage OFDM to compensate for these channel effects through pilot-based equalization. Each transmitted frame begins with a Schmidl-Cox preamble for carrier frequency offset and Doppler correction, followed by OFDM symbols carrying the learned waveform samples with pilot insertion at period~4 (75\% spectral efficiency) and cyclic prefix for inter-symbol interference mitigation.}

\newtext{A key contribution of E2E-WAVE is rendering this entire transmission pipeline differentiable to enable end-to-end optimization from video reconstruction loss to waveform parameters. Standard OFDM implementations use hard indexing for resampling operations (upsampling via cyclic prefix insertion and pulse shaping, downsampling via matched filtering and symbol extraction), which blocks gradient flow. We instead implement all resampling through linear interpolation, computing weighted combinations of adjacent samples for upsampling and employing fractional indexing with linear interpolation for downsampling at non-integer positions. Beyond differentiable resampling, we must also enable backpropagation through the channel simulation itself. We employ realistic underwater acoustic channel models from at-sea measurements (Watermark dataset~\cite{van2017watermark}), each characterized by a time-varying impulse response matrix. The channel replay process, baseband conversion, polyphase resampling to match channel sampling rate, convolution with impulse response, polyphase resampling back to transmission rate, and passband conversion, is implemented entirely through differentiable operations by representing convolutions as matrix multiplications and resampling as polyphase filters with linear interpolation. This full-stack differentiability allows the waveform bank to learn representations adapted to the statistical characteristics of underwater acoustic propagation.}

\newtext{During training, video frames are encoded to tokens via the frozen VideoGPT encoder, mapped to waveforms through the trainable bank, modulated with OFDM (preamble and pilots), and transmitted through simulated channels applying frequency-selective fading, multipath, and AWGN at specified SNR levels. The receiver performs Schmidl-Cox carrier frequency offset compensation and synchronization, pilot-based channel estimation, zero-forcing equalization per subcarrier, waveform decoding via nearest-neighbor search in the learned bank, and VideoGPT reconstruction. The cross-entropy loss computed against the semantic relevance matrix backpropagates through this differentiable pipeline, jointly optimizing the waveform parameters to embed semantic similarity while accounting for realistic channel impairments.}

\section{Performance Evaluation}\label{sec:perfeval}
We evaluate E2E-WAVE against traditional coding approaches across realistic underwater acoustic channels.

\subsection{Experimental Setup}

\textbf{Video Datasets}:
We employ a train-test split using diverse video datasets to evaluate generalization to unseen content.

\textit{Training}:
DRUVA~\cite{nishavarghese2022druva} provides 15 hours of footage at 50--300m depths with marine biodiversity scenarios, UOT100~\cite{kezebou2023uot100} contributes 96,000 frames across 104 sequences covering 10 underwater object categories, and UCF101~\cite{soomro2012ucf101} adds 13,320 videos spanning 101 human action classes to improve generalization across diverse visual content. The underwater datasets span the full spectrum of visibility conditions---from clear water with vivid details to turbid scenes with heavy particulates, as illustrated in Fig.~\ref{fig:dataset_examples}: Top row.

\begin{table}[t]
\centering
\caption{OFDM Transmission Parameters}
\label{tab:ofdm_params}
\begin{tabular}{lr}
\toprule
\textbf{Parameter} & \textbf{Value} \\
\midrule
Subcarriers & 64 \\
Cyclic prefix & 63 samples \\
Symbols per frame & 16 \\
Pilot period & 4 \\
Bandwidth & 8~kHz \\
Frame duration & 336~ms \\
Chirp Sync Head & 500 samples\\
Schmidl-Cox & 128 samples \\
\bottomrule
\end{tabular}
\vspace{-0.4cm}
\end{table}

\textit{Testing}:
We evaluate exclusively on the UVE-38K dataset~\cite{qi2021underwater}, entirely unseen during training, comprising 38,000+ frames across 50 videos in 7 scene categories. We partition the test set into two categories: \textit{turbid} (33 videos) and \textit{clear} (11 videos), as illustrated in Fig.~\ref{fig:dataset_examples}: Bottom row. Under extreme compression at sub-5~kbps rates, turbid videos---characterized by low visibility, diffuse lighting, and reduced scene complexity---yield higher PSNR/SSIM as their limited information content can be adequately captured by the few tokens we can transmit. Conversely, clear videos with vivid colors, sharp coral textures, and caustic light patterns contain substantially more information, resulting in lower reconstruction metrics when constrained to the same token budget.

\begin{figure*}[t]
	\centering
	\includegraphics[width=\textwidth]{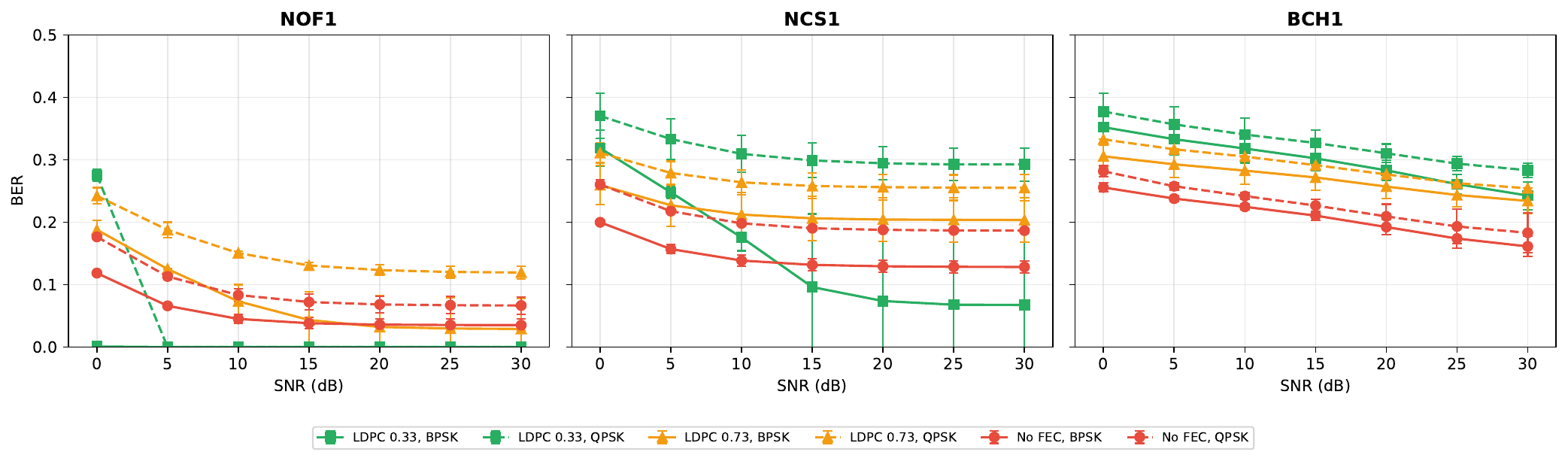}
	\caption{Bit error rate (BER) vs. SNR across Watermark channel environments for BPSK and QPSK modulation. Even the most favorable channel (NOF1) exhibits 8.6\% BER with QPSK at operational SNR, while harsher channels (BCH1, KAU1) reach 20--46\% BER, rendering traditional video codecs unusable.}
	\label{fig:ber_comparison}
\end{figure*}

\textbf{Underwater Channel Datasets and Simulation}:
For realistic channel characterization, we employ the Watermark benchmark dataset~\cite{socheleau2016watermark}, which provides Time-Varying Impulse Responses~(TVIRs) from at-sea measurements across distinct environments with delay spreads of 100--130~ms. Critically, \textbf{E2E-WAVE is trained exclusively on the NCS1 channel} and evaluated on three channels: NCS1 (training channel), NOF1, and BCH1. The latter two represent geographically and acoustically distinct environments never seen during training, providing a rigorous test of cross-location generalization.

The three evaluation channels exhibit diverse acoustic characteristics. \textbf{NCS1} (Continental Shelf, Norway) is a Single-Input Single-Output~(SISO) channel with 60 recordings spanning 33 minutes total playtime, operating at 8~kHz bandwidth with 14~kHz carrier frequency. \textbf{NOF1} (Oslofjord, Norway) shares the same SISO configuration with 60 recordings (33 min total), 8~kHz bandwidth, and 14~kHz carrier, but represents a geographically distinct acoustic environment. \textbf{BCH1} (Brest Commercial Harbor, France) is a Single-Input Multiple-Output~(SIMO) channel with 4 recordings (one per hydrophone) providing 1~minute playtime per hydrophone, operating at 10~kHz bandwidth with 35~kHz carrier frequency. Importantly, we leverage BCH1's SIMO configuration to evaluate \textbf{multicasting} performance by transmitting the same video frame through all 4 hydrophone recordings. This simulates a realistic multicast scenario where multiple underwater nodes at different locations and SNR conditions simultaneously receive the same transmission. Performance on BCH1 thus directly demonstrates E2E-WAVE's ability to support multicast communication without requiring additional adaptations.

Table~\ref{tab:ofdm_params} summarizes OFDM parameters. Each frame begins with a chirp for synchronization and Schmidl-Cox preamble for Carrier Frequency Offset~(CFO) estimation, followed by OFDM symbols. Pilots inserted every fourth symbol enable frequency-selective channel estimation; zero-forcing equalization then compensates multipath per subcarrier. A cyclic prefix provides the guard interval. Furthermore, the power is normalized to 1.0 to simulate real-life transmission.

\begin{figure*}[t]
	\centering
	\includegraphics[width=\textwidth]{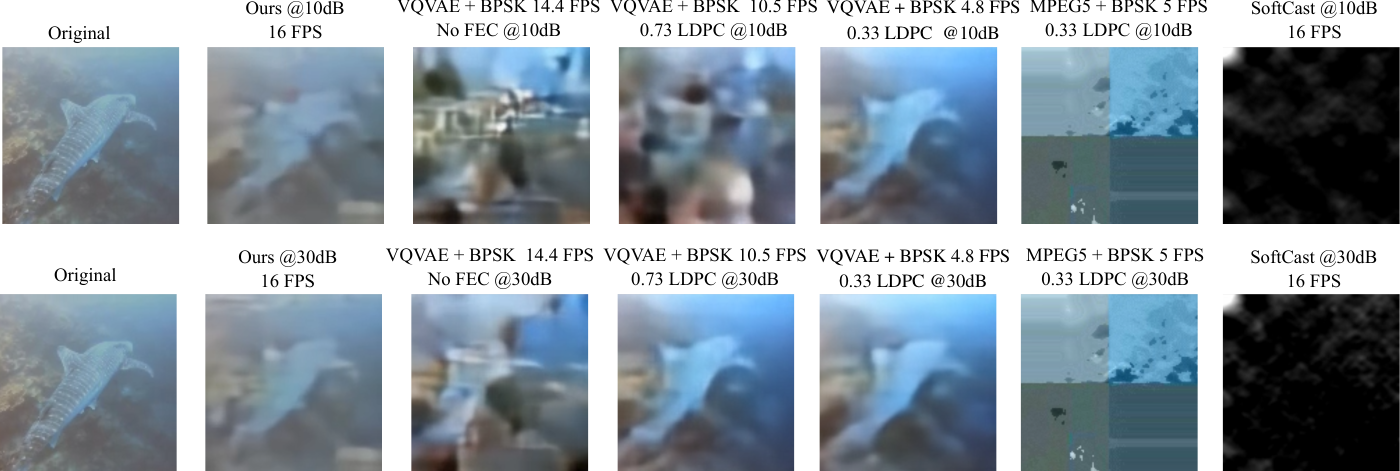}
	\caption{Qualitative reconstruction comparison on NOF1 channel at 10~dB and 30~dB SNR. \textbf{Bottom row}: At 30~dB, E2E-WAVE and VideoGPT+BPSK with LDPC $r=0.33$ both produce recognizable reconstructions, while HEVC and SoftCast fail completely. \textbf{Top row}: At 10~dB, E2E-WAVE maintains coherent output while VideoGPT+BPSK with LDPC $r=0.73$ degrades significantly. HEVC fails at all SNR levels due to insufficient bitrate after FEC overhead; SoftCast fails due to frequency-selective fading violating its AWGN assumption.}
	\label{fig:qual_comparison}
\end{figure*}

The available throughput is severely limited. With an 8~kHz bandwidth and 64 subcarriers, each OFDM symbol spans $127/8000 = 15.9$~ms (including CP). A 336~ms frame contains 16 OFDM symbols, of which 4 are pilots, leaving 12 data symbols $\times$ 64 subcarriers $= 768$ data symbols per frame. This yields a raw throughput of $768/0.336 \approx 2.3$~kbps for BPSK and 4.6~kbps for QPSK---before any error correction.
The severity of these channels is evident from the raw bit error rates~(BER) measured \emph{before and after} Forward Error Correction~(FEC), as shown in Fig.~\ref{fig:ber_comparison}. The best channel~(NOF1) achieves 8.6\% BER with QPSK, while the harsher channels exhibit 20--46\% BER. At these error rates, traditional digital transmission is fundamentally unreliable. HEVC and H.264 require near-perfect bit delivery---a single corrupted byte can desynchronize the decoder, causing catastrophic frame loss. Moreover, H.264's minimum target bitrate is 1~kbps; it simply cannot encode video at sub-1~kbps rates, leaving us limited to H.265 codec. Furthermore, as observed in Fig.~\ref{fig:ber_comparison}, a curious phenomenon occurs on the harshest channels (NCS1, BCH1): when raw BER exceeds the FEC's decoding threshold, the iterative decoder fails to converge and can output \textit{worse} errors than uncoded transmission---a well-documented cliff effect in coding theory~\cite{richardson2001capacity}. We observed LDPC $r=0.73$ increasing BER from 13--15\% (uncoded) to 22--25\% (coded) on these channels, rendering conventional FEC counterproductive.

\begin{table*}[t]
	\centering
	\small
	\caption{Fair comparison configurations between digital modulation and E2E-WAVE at equivalent throughput. With a 1024-token codebook, digital transmission requires 10 bits per token; the wavelength $L$ denotes OFDM symbols per token for E2E-WAVE. BPSK columns show bits (digital) or symbols (E2E-WAVE) per token; QPSK halves these requirements. FEC overhead reduces achievable FPS proportionally. Notably, real-time 16 FPS is \textit{impossible} for digital modulation ($2^9=512 < 1024$ tokens) but achievable with E2E-WAVE at $L=9$.}
	\label{tab:wavebank_equivalence}
	\begin{tabular}{lccccc}
		\toprule
		\textbf{Digital Config.} & \textbf{FPS} & \textbf{BPSK} & \textbf{QPSK} & \textbf{E2E-WAVE (BPSK Equiv.)} & \textbf{E2E-WAVE (QPSK Equiv.)} \\
		\midrule
		Real-time & 16.0 & -- & -- & $L=9$ & $L=5$ \\
		No FEC & 14.4 & 10 bits & 5 sym & $L=10$ & $L=5$ \\
		LDPC $r=0.73$ & 10.5 & 14 bits & 7 sym & $L=13$ & $L=7$ \\
		LDPC $r=0.33$ & 4.8 & 30 bits & 15 sym & $L=30$ & $L=15$ \\
		\bottomrule
	\end{tabular}
\end{table*}

\begin{figure*}[t]
	\centering
	\includegraphics[width=\textwidth]{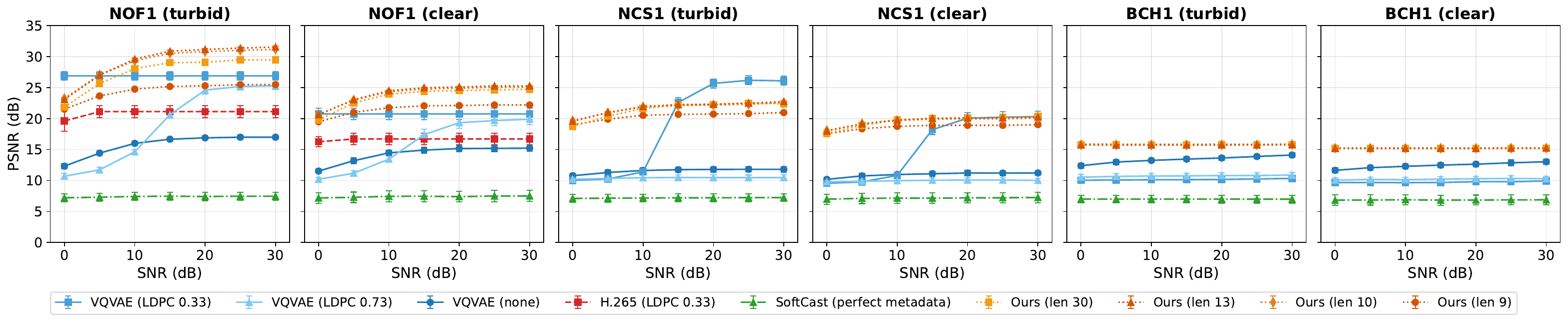}
	\caption{PSNR vs SNR comparison across baselines for BPSK modulation on UVE-38K test set. Each subplot shows performance on a specific Watermark channel (NOF1, NCS1, BCH1) and scene type (clear or turbid). \textbf{E2E-WAVE is trained only on NCS1}; performance on NOF1 and BCH1 demonstrates cross-location generalization to unseen channel conditions. Legend entries ``Ours (len $x$)'' refer to E2E-WAVE wavebank wavelengths from Table~\ref{tab:wavebank_equivalence}. Turbid scenes achieve higher PSNR due to lower visual complexity, while clear scenes exhibit lower PSNR due to higher information content.}
	\label{fig:psnr_vs_snr_bpsk}
    \vspace{-0.4cm}
\end{figure*}

\begin{figure*}[t]
	\centering
	\includegraphics[width=\textwidth]{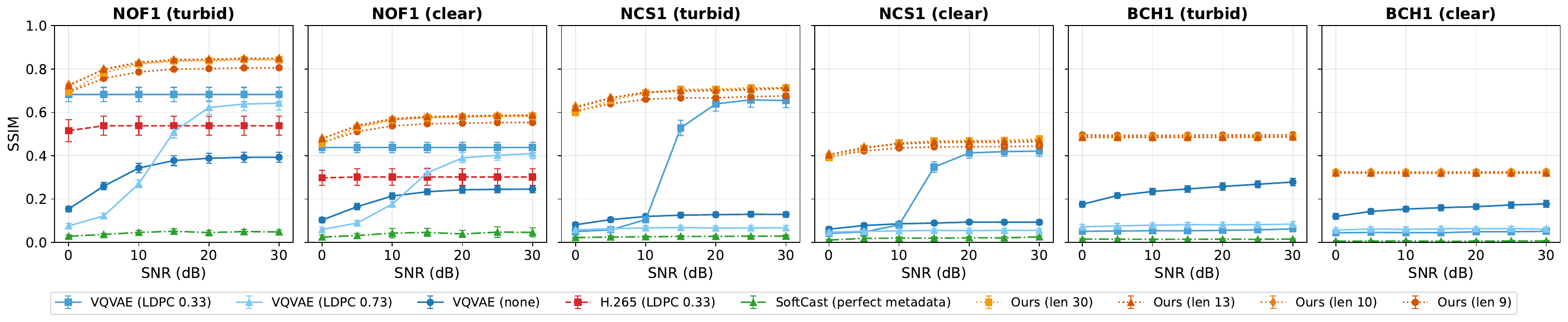}
	\caption{SSIM vs SNR comparison across baselines for BPSK modulation on UVE-38K test set. Each subplot shows performance on a specific Watermark channel (NOF1, NCS1, BCH1) and scene type (clear or turbid). \textbf{E2E-WAVE is trained only on NCS1}; performance on NOF1 and BCH1 demonstrates cross-location generalization to unseen channel conditions. Legend entries ``Ours (len $x$)'' refer to E2E-WAVE wavebank wavelengths from Table~\ref{tab:wavebank_equivalence}. Structural similarity follows similar trends to PSNR, with E2E-WAVE demonstrating graceful degradation across all channel conditions.}
	\label{fig:ssim_vs_snr_bpsk}
    \vspace{-0.4cm}
\end{figure*}

Analog schemes like SoftCast face equally severe limitations. With effective rates dropping below 1~kbps after FEC overhead, SoftCast's linear scaling of quality with SNR provides little benefit when the available bandwidth cannot support even a single low-resolution frame per second. Furthermore, SoftCast assumes an AWGN channel; in frequency-selective UWA channels, zero-forcing equalization amplifies noise at deep fades and fails to fully compensate for multipath distortion, degrading reconstruction quality. These constraints motivate our learned waveform approach, which exploits semantic similarity between tokens to achieve graceful degradation without requiring perfect bit recovery.

\textbf{Baselines}:
We compare E2E-WAVE against three baselines spanning learned video tokenizers with traditional modulation, analog transmission, and conventional codecs. All methods are evaluated separately on \textbf{clear} (11 videos) and \textbf{turbid} (33 videos) scenes from the UVE-38K test set:

\begin{itemize}
    \item \textbf{VideoGPT VQVAE + BPSK/QPSK}: VideoGPT VQVAE tokenizer~\cite{yan2021videogpt} with traditional digital modulation, evaluated with and without FEC.
    \item \textbf{SoftCast}: Analog joint source-channel coding~\cite{jakubczak2010softcast} with perfect metadata protection.
    \item \textbf{HEVC (H.265)}: Standard video codec at maximum bitrate for 16 fps, transmitted via BPSK/QPSK with FEC protection.
    \item \textbf{E2E-WAVE (Ours)}: VideoGPT VQVAE encoder with learned wavebank, no FEC required.
\end{itemize}
This comparison isolates the contribution of our wavebank approach by using the same VideoGPT VQVAE tokenizer as one of the baselines, differing only in the transmission method (learned wavebank vs. traditional BPSK/QPSK modulation).

\textbf{Hardware Used}:
All experiments are conducted on a Dell Precision 7920 Tower Workstation equipped with dual Intel Xeon processors, 128 GB RAM, and two NVIDIA RTX 2080 Ti GPUs with 11 GB VRAM each.

\subsection{Evaluation Metrics}
Our E2E-WAVE evaluation pipeline assesses performance using the peak signal-to-noise ratio~(PSNR) and structural similarity~(SSIM) to quantify reconstruction quality.

\textbf{Qualitative Results}:
Fig.~\ref{fig:qual_comparison} presents visual reconstruction examples on the NOF1 channel---the most favorable of our test environments. Even on this best-case channel, the limitations of conventional approaches become apparent. At high SNR (30~dB), only VideoGPT+BPSK with heavy FEC (LDPC $r=0.33$) approaches E2E-WAVE quality, but at the cost of reducing frame rate to 4.8~fps. At moderate SNR (10~dB), the performance gap widens: VideoGPT+BPSK with lighter FEC (LDPC $r=0.73$) produces visibly degraded output, while E2E-WAVE maintains perceptually coherent reconstructions. HEVC and SoftCast fail across all conditions---HEVC because sub-5~kbps bandwidth cannot accommodate both video data and necessary FEC overhead, SoftCast because its linear analog transmission amplifies noise at frequency-selective fades rather than gracefully degrading.

\begin{figure*}[t]
	\centering
	\includegraphics[width=\textwidth]{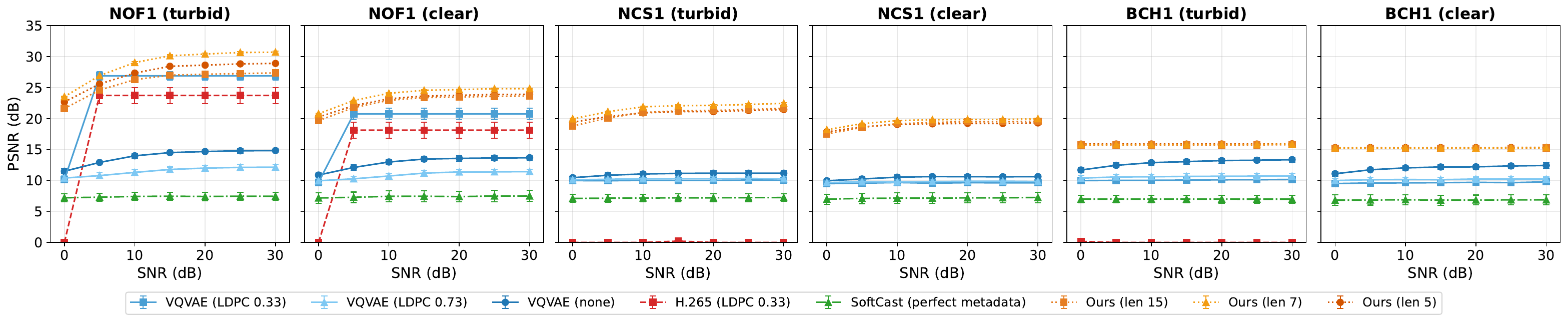}
	\caption{PSNR vs SNR comparison across baselines for QPSK modulation on UVE-38K test set. Each subplot shows performance on a specific Watermark channel (NOF1, NCS1, BCH1) and scene type (clear or turbid). \textbf{E2E-WAVE is trained only on NCS1}; performance on NOF1 and BCH1 demonstrates cross-location generalization to unseen channel conditions. Legend entries ``Ours (len $x$)'' refer to E2E-WAVE wavebank wavelengths from Table~\ref{tab:wavebank_equivalence}. QPSK doubles the throughput compared to BPSK but exhibits higher sensitivity to channel impairments.}
	\label{fig:psnr_vs_snr_qpsk}
    \vspace{-0.4cm}
\end{figure*}

\begin{figure*}[t]
	\centering
	\includegraphics[width=\textwidth]{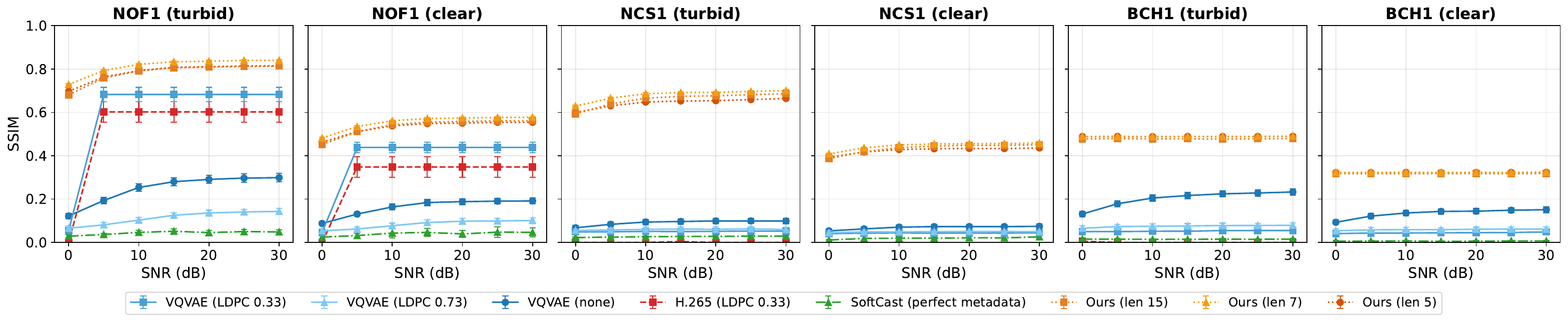}
	\caption{SSIM vs SNR comparison across baselines for QPSK modulation on UVE-38K test set. Each subplot shows performance on a specific Watermark channel (NOF1, NCS1, BCH1) and scene type (clear or turbid). \textbf{E2E-WAVE is trained only on NCS1}; performance on NOF1 and BCH1 demonstrates cross-location generalization to unseen channel conditions. Legend entries ``Ours (len $x$)'' refer to E2E-WAVE wavebank wavelengths from Table~\ref{tab:wavebank_equivalence}. Structural similarity trends mirror PSNR results, with E2E-WAVE maintaining robust performance across channel conditions.}
	\label{fig:ssim_vs_snr_qpsk}
    \vspace{-0.4cm}
\end{figure*}

\textbf{Wavebank Configuration and Fair Comparison}:
The critical parameter for our wavebank is the \textit{wavelength}---the number of data symbols in an OFDM frame used to encode each token index. To establish a fair comparison with traditional digital modulation, we derive equivalent configurations based on throughput constraints.
With a VideoGPT VQVAE codebook of size 1024, each token requires $\log_2(1024) = 10$ bits for digital transmission. Our $128 \times 128$ video with $4 \times 16 \times 16$ compression yields 16 tokens per frame. At 16~fps, real-time transmission demands $16 \times 10 \times 16 = 2560$ bits per second. However, as established based on data presented in Table~\ref{tab:ofdm_params}, our channel supports only 2310~bps with BPSK---insufficient for real-time video. Raw BPSK without FEC can therefore deliver at most $2310/160 \approx 14.43$ equivalent fps.

For fair comparison, our wavebank with wavelength $L=10$ (10 symbols per token) operates at the same rate as uncoded BPSK, achieving 14.43~fps. When baselines employ FEC, we correspondingly increase the wavebank wavelength: $L=13$ matches LDPC rate $r=0.73$ (10.53~fps), and $L=30$ matches LDPC rate $r=0.33$ (4.76~fps). Table~\ref{tab:wavebank_equivalence} summarizes these equivalences for both BPSK and QPSK modulation. Crucially, our learned wavebank is not constrained to integer bit boundaries---with $L=9$ symbols per token, we achieve the full 16~fps target, a configuration \textit{impossible} for digital modulation since $2^9 = 512 < 1024$ tokens. This flexibility is a unique advantage of the learned waveform representations.

\begin{figure*}[t]
	\centering
	\includegraphics[width=\textwidth]{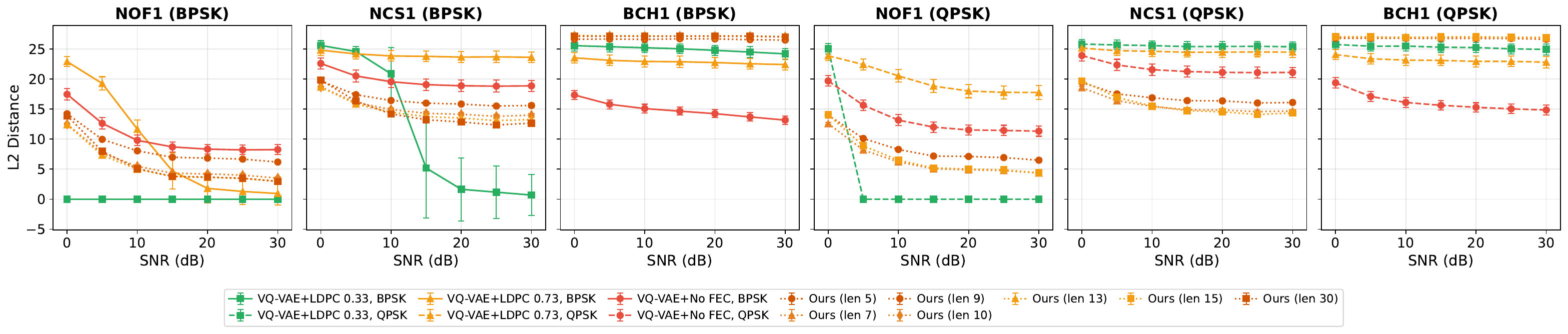}
	\caption{Average $L_2$ distance between transmitted and received token embeddings vs. SNR. Traditional digital modulation (BPSK/QPSK with FEC) treats all bit errors equally, corrupting tokens indiscriminately regardless of semantic content---high $L_2$ distance directly translates to poor reconstruction. E2E-WAVE's learned wavebank exploits inter-token semantic relevance: even when channel degradation increases $L_2$ distance, errors preferentially map to semantically similar tokens, maintaining reconstruction quality.}
	\label{fig:l2_comparison}
    \vspace{-0.4cm}
\end{figure*}

\textbf{Quantitative Results}:
Fig.~\ref{fig:psnr_vs_snr_bpsk} and Fig.~\ref{fig:ssim_vs_snr_bpsk} present PSNR and SSIM performance for BPSK, while Fig.~\ref{fig:psnr_vs_snr_qpsk} and Fig.~\ref{fig:ssim_vs_snr_qpsk} show QPSK results. E2E-WAVE is trained exclusively on NCS1; performance on NOF1 and BCH1 demonstrates cross-location generalization. On NOF1 at 30~dB SNR, E2E-WAVE achieves 30~dB PSNR and 0.82 SSIM at $L=30$, compared to 27~dB PSNR and 0.70 SSIM for VideoGPT+BPSK with LDPC $r=0.33$---a +3~dB PSNR and +0.12 SSIM gain at equivalent throughput. Notably, E2E-WAVE maintains robust performance across a wide range of wavelengths ($L=10$ to $L=30$), with $L=15$ achieving 27~dB PSNR and 0.80 SSIM---demonstrating that our learned wavebank technique can deliver performance across different throughput configurations. H.265 reaches only 21~dB PSNR (0.55 SSIM), while SoftCast fails catastrophically at 7~dB PSNR. On harsher channels, the gap widens: E2E-WAVE ($L=30$) maintains 21~dB PSNR (0.60 SSIM) on NCS1 and 16~dB PSNR (0.50 SSIM) on BCH1, whereas baselines plateau at 27~dB/0.62 (NCS1) and 12~dB/0.25 (BCH1). With QPSK on harsh channels, baselines cannot exceed 15~dB PSNR or 0.22 SSIM, rendering higher-order modulation counterproductive.

\textbf{Graceful Degradation}:
Fig.~\ref{fig:l2_comparison} quantifies E2E-WAVE's key advantage by measuring the average $L_2$ distance between transmitted and received token embeddings across SNR levels. For traditional FEC-protected digital transmission, token corruption is indiscriminate---bit errors map transmitted tokens to arbitrary received tokens without regard for semantic similarity, and the resulting high $L_2$ distances directly degrade reconstruction quality. In contrast, E2E-WAVE's learned wavebank embeds semantic structure into the physical layer: when channel noise corrupts the received waveform, the nearest-neighbor decoder preferentially selects tokens that are semantically related to the original. This explains why E2E-WAVE maintains perceptual quality even as raw token accuracy decreases---the ``errors'' are biased toward visually similar content rather than random corruption.

\section{Conclusion and Future Work}\label{sec:concl}

We presented E2E-WAVE, the first to embed semantic similarity directly into physical-layer waveforms for underwater video transmission. Our trainable waveform bank, optimized via cross-entropy loss against codebook-derived semantic relevance, ensures channel-induced decoding errors preferentially select semantically similar tokens, achieving graceful degradation. Combined with VideoGPT tokenization (1024$\times$ compression at $\sim$24~dB PSNR), E2E-WAVE achieves 16~fps real-time transmission at 128$\times$128 resolution over severely bandwidth-constrained underwater channels, outperforming VideoGPT with digital modulation and FEC, HEVC, and SoftCast by substantial margins in +5dB (19.26\%) PSNR and +0.10 (14.28\%) SSIM on less challenging channels (NOF1), while the margin is even higher on harsher channels.

To enable end-to-end optimization, we developed a fully differentiable transmission pipeline supporting OFDM equalization through differentiable resampling (linear interpolation) and channel replay (matrix convolutions with polyphase filtering). Trained exclusively on NCS1, E2E-WAVE generalizes to unseen channels (NOF1, BCH1) without retraining, demonstrating cross-location deployability. Our BCH1 SIMO evaluation with 4 independent receivers confirms native multicasting capability without metadata or receiver-specific adaptations. The learned wavebank's flexibility enables real-time transmission at wavelength $L=9$, impossible for digital modulation with a 1024-token codebook.

\textbf{Future Directions:} \textit{(1)~Joint tokenizer-wavebank optimization}: End-to-end training of both tokenizer and waveform bank could learn codebooks optimized for underwater channels, improving semantic structure and error resilience. \textit{(2)~Hierarchical transmission for higher resolutions}: Extending to 256$\times$256+ via hierarchical tokenization---base tokens use robust long-wavelength waveforms ($L=20$--30) for coarse reconstruction, enhancement tokens use shorter wavelengths ($L=5$--10) for detail, enabling graceful quality scaling across SNRs. \textit{(3)~Online adaptation}: Mobile platforms experience dynamic Doppler and multipath; lightweight adapter networks mapping channel estimates to waveform adjustments could maintain performance without retraining. \textit{(4)~Deployment validation}: Field testing on acoustic modems will reveal hardware noise and optimization opportunities.

\balance

\bibliographystyle{ieeetr}
\bibliography{refs}

\end{document}

%% file: preamble/units.tex
\usepackage{units}


%% file: preamble/markupAndCommenting.tex
\newcommand{\newcolorlabel}[2]{%
  \expandafter\newcommand\csname #1\endcsname[1]{%
    \colorbox{#2}{\color{white}\textsf{\textbf{##1}}}}%
}

\newcommand{\newcommenter}[2]{%
  \expandafter\newcommand\csname #1\endcsname[1]{%
    \fcolorbox{#2}{#2}{\color{white}\textsf{\textbf{#1}}}
    {\color{#2}##1}}%
  \expandafter\newcommand\csname at#1\endcsname{%
    \fcolorbox{#2}{#2}{\color{white}\textsf{\textbf{@#1}}}
    {\color{#2}}}%
  \expandafter\newcommand\csname #1hl\endcsname[2]{%
    \colorbox{#2}{\color{white}\textsf{\textbf{#1}}}\fcolorbox{#2}{white}{##2}
    {\color{#2}##1}}%
}

\newcommenter{TODO}{DodgerBlue1}
\newcommenter{rtron}{Green3}

\usepackage{comment}

\usepackage{pdfcomment}

\usepackage{soul}

\usepackage[normalem]{ulem}

\usepackage{csquotes}
